\documentclass[prd,aps,showpacs,,nofootinbib,superscriptaddress,preprintnumbers,epsf,psf]{revtex4}
\usepackage[english]{babel}
\usepackage{pstricks}
\usepackage[dvips]{graphicx}
\usepackage{epsf}

\newcommand{\qq}{Q'} 
\newcommand{\re}{\mathrm{Re}\,}
\newcommand{\im}{\mathrm{Im}\,}
\newcommand{\gev}{\,{\rm GeV}}
\newcommand{\tev}{\,{\rm TeV}}
\def\pb{\,{\rm pb}}

\newcommand{\AmS}{{\protect\the\textfont2
  A\kern-.1667em\lower.5ex\hbox{M}\kern-.125emS}}

\begin{document}
 \makeatletter
 \def\preprint#1{
 }
\preprint{\begin{tabular}{l}
      arXiv \\
      CPHT-RR080.1008
    \end{tabular}
 }
%
%
\hfill CPHT-RR080.1008
\title{Can one measure timelike Compton scattering at LHC ?}
\author{B. Pire}
 \address{ CPHT, {\'E}cole Polytechnique, CNRS, 91128 Palaiseau, France}
  \author{     L. Szymanowski 
        and
J. Wagner}
 \address{ Soltan Institute for Nuclear Studies, Ho\.{z}a 69, 00-681
Warsaw, Poland}

\begin{abstract}
Exclusive photoproduction of dileptons, $\gamma N\to
\ell^+\!\ell^- \,N$, is and will be measured in ultraperipheral
collisions at hadron colliders, such as the Tevatron, RHIC and the LHC .
We demonstrate that   the timelike deeply virtual Compton scattering (TCS)
mechanism  $\gamma q \to \ell^+\!\ell^- q $ where
the lepton pair comes from the subprocess      $\gamma q \to \gamma^* q $
dominates in some accessible kinematical regions, thus opening a new
  way  to study generalized parton distributions (GPD)    in     the nucleon.
High energy kinematics enables to probe parton distributions at
small skewedness. This subprocess interferes at the amplitude level with
the pure QED subprocess $\gamma \gamma^* \to
\ell^+\!\ell^- $ where the virtual photon is radiated from the nucleon.

\end{abstract}
%

\maketitle
\noindent
\section{ Introduction.}

Much theoretical and experimental progress  has recently
been witnessed in  the study of deeply virtual Compton scattering (DVCS),
 i.e., $\gamma^* p \to \gamma p$, 
an exclusive reaction where generalized parton
distributions (GPDs) factorize from perturbatively calculable coefficient functions, when
the virtuality of the incoming photon is high enough~\cite{Muller:1994fv}.
It is now recognized that the measurement of GPDs should contribute in a decisive way to
our understanding of how quarks and gluons build 
hadrons~\cite{gpdrev}. In particular the transverse
location of quarks and gluons become experimentally measurable via the transverse momentum dependence of the GPDs \cite{Burk}.

 The ``inverse'' process, 
 $$\gamma(q) N(p) \to \gamma^*(q') N(p') \to l^-(k) l^+(k') N(p')$$
   at small $t = (p'-p)^2$ and large \emph{timelike} virtuality $(k+k')^2=q'^2 = Q'^2$ of the final state
 dilepton, timelike Compton scattering (TCS) \cite{TCS},
shares many features with DVCS. The Bjorken variable in that case is $\tau = Q'^2/s $
 with $s=(p+q)^2$. One also defines $\Delta = p' -p$  ($t= \Delta^2$) and the skewness variables 
  $\xi\;,\;\eta$ as
\begin{eqnarray}
&&\hspace*{-0.3cm}\xi  = - \frac{(q+q')^2}{2(p+p')\cdot (q+q')} \,\approx\,
           \frac{ - Q'^2}{2s  - Q'^2} , \nonumber \\
&&\hspace*{-0.3cm}\eta = - \frac{(q-q')\cdot (q+q')}{(p+p')\cdot (q+q')} \,\approx\,
           \frac{ Q'^2}{2s  - Q'^2} ,
\label{xi-eta-def}
\end{eqnarray}
where the approximations hold in the kinematical limit we are working, i.e.
in the extended Bjorken regime
where masses and $-t$ are small with respect of $Q'^2$ ($s$ is always
larger than $Q'^2$ ).
 $x$, $\xi$, and $\eta$ represent plus-momentum fractions (Light-cone coordinates are defined as $v^\pm = \frac{v^0\pm v^3}{\sqrt{2}}$ , both proton momenta
$p$ and $p'$ moving fast to the right, i.e., having large plus-components).
\begin{equation}
x = \frac{(k+k')^+}{(p+p')^+} , \;
\xi \approx - \frac{(q+q')^+}{(p+p')^+} ,\; 
\eta \approx  \frac{(p-p')^+}{(p+p')^+} .
\nonumber
\end{equation}
To leading-twist accuracy one has  $\xi = - \eta = - \tau / (2-\tau)$.

The possibility to use high energy hadron colliders 
as powerful sources of quasi real photons in ultraperipheral collisions has
recently been emphasized \cite{UPC}. This should allow the study of many aspects
of photon proton and photon photon collisions at high energies, already
at the Tevatron and at RHIC but in particular at the
LHC \cite{UPCLHC}    even if the nominal luminosity is not achieved during
its first years of operation. The high luminosity and energies of these
photon beams opens a new kinematical domain for the study of TCS , and
thus to the hope of determining GPDs in the small skewedness ($\xi$) region, which is  complementary  to the determination
 of the large $\xi$ quark  GPDs at lower energy electron accelerators such as
JLab. Moreover, the crossing from 
 a spacelike to a timelike probe is an important test of the understanding of QCD 
 corrections, as shown by the history of the understanding of the Drell-Yan reaction
  in terms of QCD.

\begin{figure}[htb]
\begin{center}
     \epsfxsize=0.5\textwidth
     \epsffile{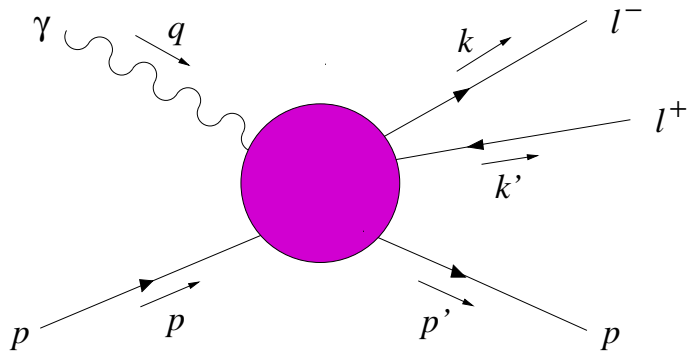}
\caption{\label{refig}Real photon-proton scattering into a lepton pair
and a proton.}
\end{center}
\end{figure}
%

The physical process where to observe TCS is photoproduction of a
heavy lepton pair, $\gamma N \to \mu^+\!\mu^-\, N$ or $\gamma N \to
e^+\!e^-\, N$, shown in Fig.~\ref{refig}. As in the case of DVCS, a Bethe-Heitler (BH)
mechanism - sometimes called $\gamma \gamma $ process since the lepton pair is produced through the 
$\gamma (q)\gamma (\Delta) \to\ell^+\ell^-$ subprocess -  contributes at the amplitude level. This amplitude 
is completely calculable in QED provided one knows the  Nucleon form factors at small $t$.
This process has a very peculiar angular dependence and overdominates the TCS process if
one blindly integrates over the final phase space. One may however choose kinematics where 
the amplitudes of the two processes are of the same order of magnitude, and either subtract the 
well-known Bethe-Heitler process or use specific observables sensitive to
 the interference of the two amplitudes.

\begin{figure}[tb]
\begin{center}
  \epsfxsize=0.95\textwidth
  \epsffile{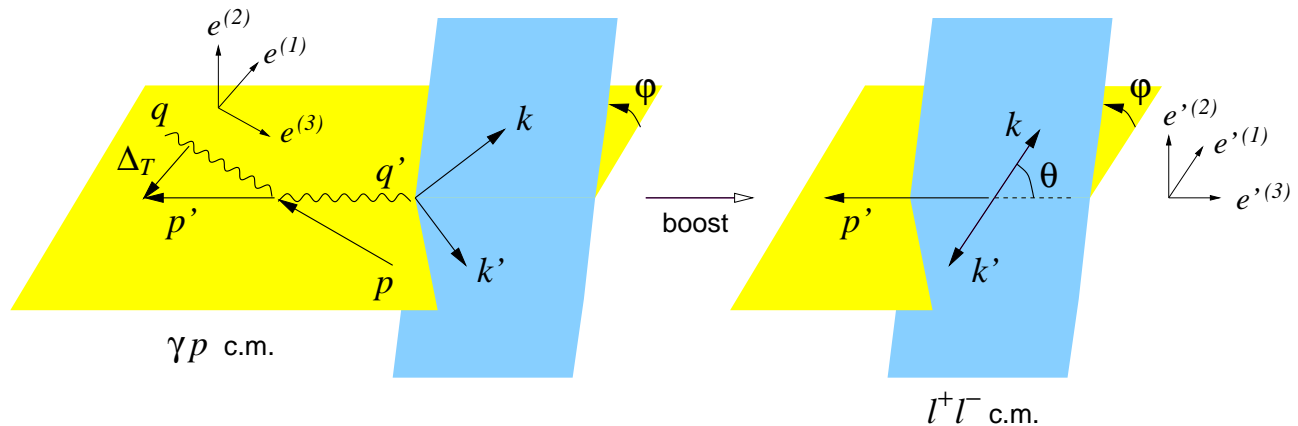}
\caption{Kinematical variables and coordinate axes in
the $\gamma p$ and $\ell^+\ell^-$ c.m.\ frames. }
\label{angle}
\end{center}
\end{figure}

The kinematics of the $\gamma (q) N (p)\to \ell^-(k) \ell^+(k') N(p')$ process is shown in Fig. \ref{angle}.
 In the $\ell^+\ell^-$ center of mass system,  one introduces the polar and azimuthal angles $\theta$
and $\varphi$ of $\vec{k}$, with reference to a coordinate system with
$3$-axis along $-\vec{p}\,'$ and $1$- and $2$-axes such that $\vec{p}$
lies in the $1$-$3$ plane and has a positive
$1$-component.

%

In this paper, we shall examine in detail the feasibility of  TCS experiments in ultraperipheral
 collisions at the LHC. Most
conclusions should also apply to the Tevatron and RHIC experimental
conditions, once the effective photon fluxes and energies are scaled down
to their specific values. 
We shall work in the leading twist approximation. As in the DVCS case,  a gauge invariant 
treatment necessitates the inclusion of twist 3 effects \cite{APT}, but we will not address this
problem in this paper. We will also stay 
at the leading order in $\alpha_{S}$, and thus neglect  contributions proportionnal 
to gluon GPDs. The motivation of this study has been presented at a recent LHC workshop \cite{PSW}.

\section{The various contributions}
\subsection{The Bethe-Heitler contribution}
\label{sec:bethe}

\begin{figure}[tb]
\begin{center}
     \epsfxsize=0.8\textwidth
      \epsffile{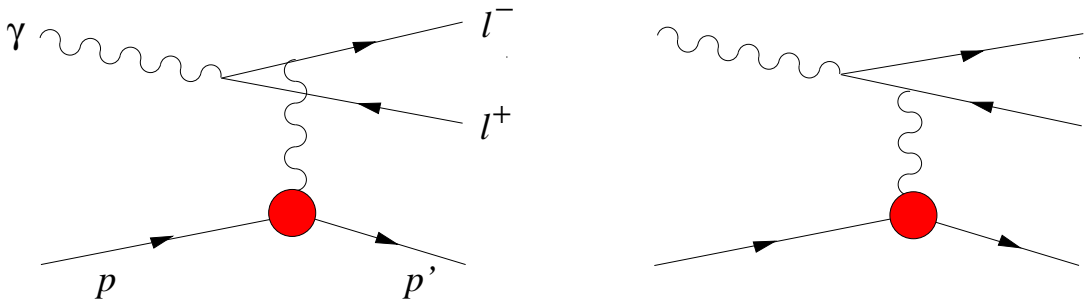}
\caption{The Feynman diagrams for the Bethe-Heitler amplitude.}
\label{bhfig}
\end{center}
\end{figure}

The Bethe-Heitler amplitude is  calculated from the two Feynman
diagrams in Fig.~\ref{bhfig}
 where the photon-nucleon vertex is
parameterized by the usual Dirac 
 and Pauli form factors $F_1(t)$ and
$F_2(t)$, normalizing $F_2(0)$ to be the anomalous magnetic moment of
the target.  Neglecting masses and $t$  compared to
terms going with $s$ or $\qq ^2$, the  Bethe Heitler contribution to the unpolarized
$\gamma p$ cross section is ($M$ is the proton mass) 
\begin{eqnarray}
  \label{approx-BH}
\frac{d \sigma_{BH}}{d\qq ^2\, dt\, d(\cos\theta)\, d\varphi}
  \approx
    \frac{\alpha^3_{em}}{2\pi s^2}\, \frac{1}{-t}\,
    \frac{1 + \cos^2\theta}{\sin^2\theta} \,
\left[ \Big(F_1^2 -\frac{t}{4M^2} F_2^2\Big)
            \frac{2}{\tau^2}\, \frac{\Delta_T^2}{-t}\,
        + (F_1+F_2)^2 \,\right] ,
\end{eqnarray}
provided we stay away from the kinematical region where the  product  of lepton propagators goes 
to zero at very small $\theta$. The interesting physics program thus imposes a
cut on $\theta$ to stay away from the region where the Bethe Heitler  cross section becomes
extremely large.

\subsection{The Compton amplitude}
\label{sec:compton}

\begin{figure}
\begin{center}
    \epsfxsize=0.39\textwidth
     \epsffile{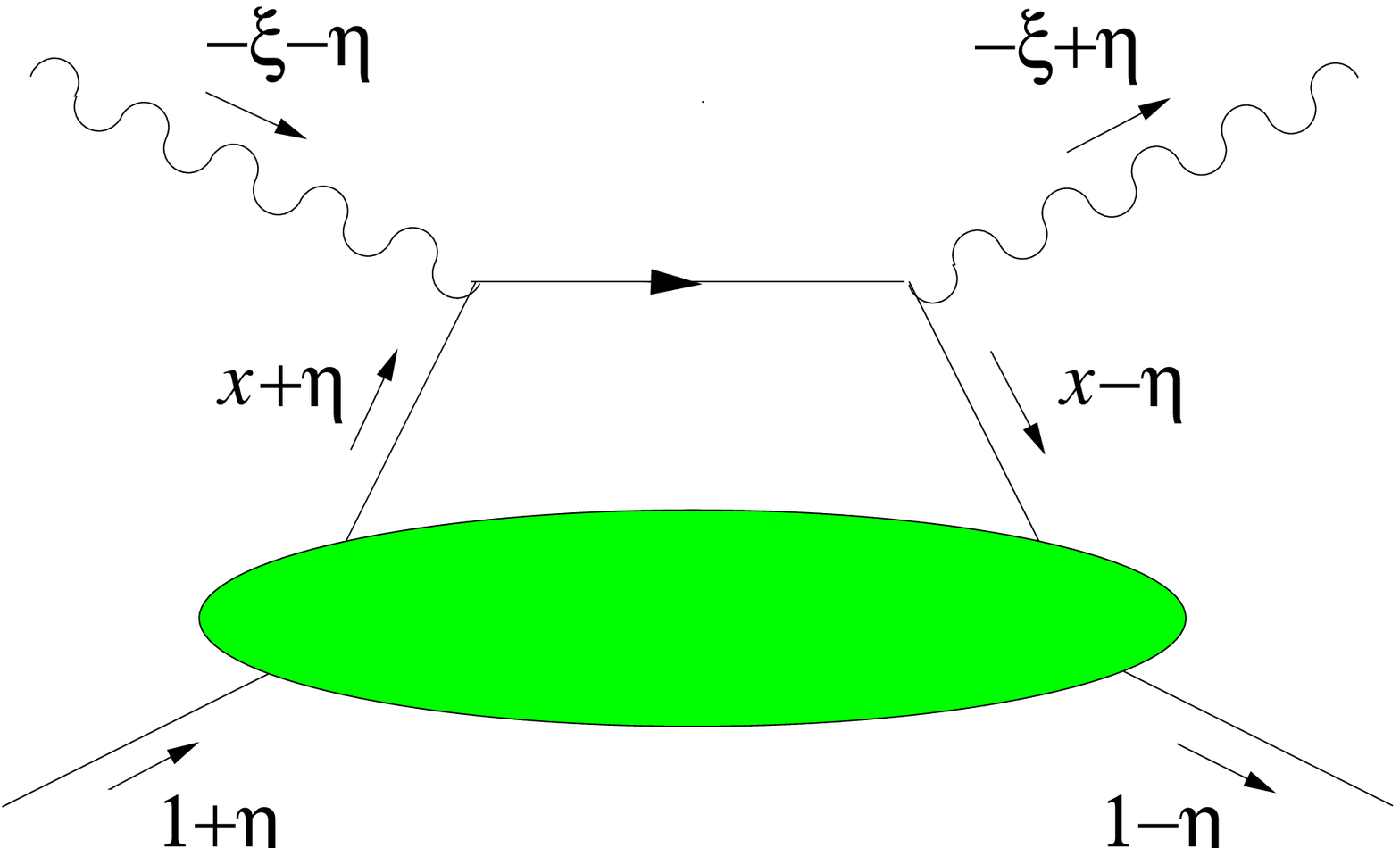}
\hspace{0.05\textwidth}
    \epsfxsize=0.39\textwidth
    \epsffile{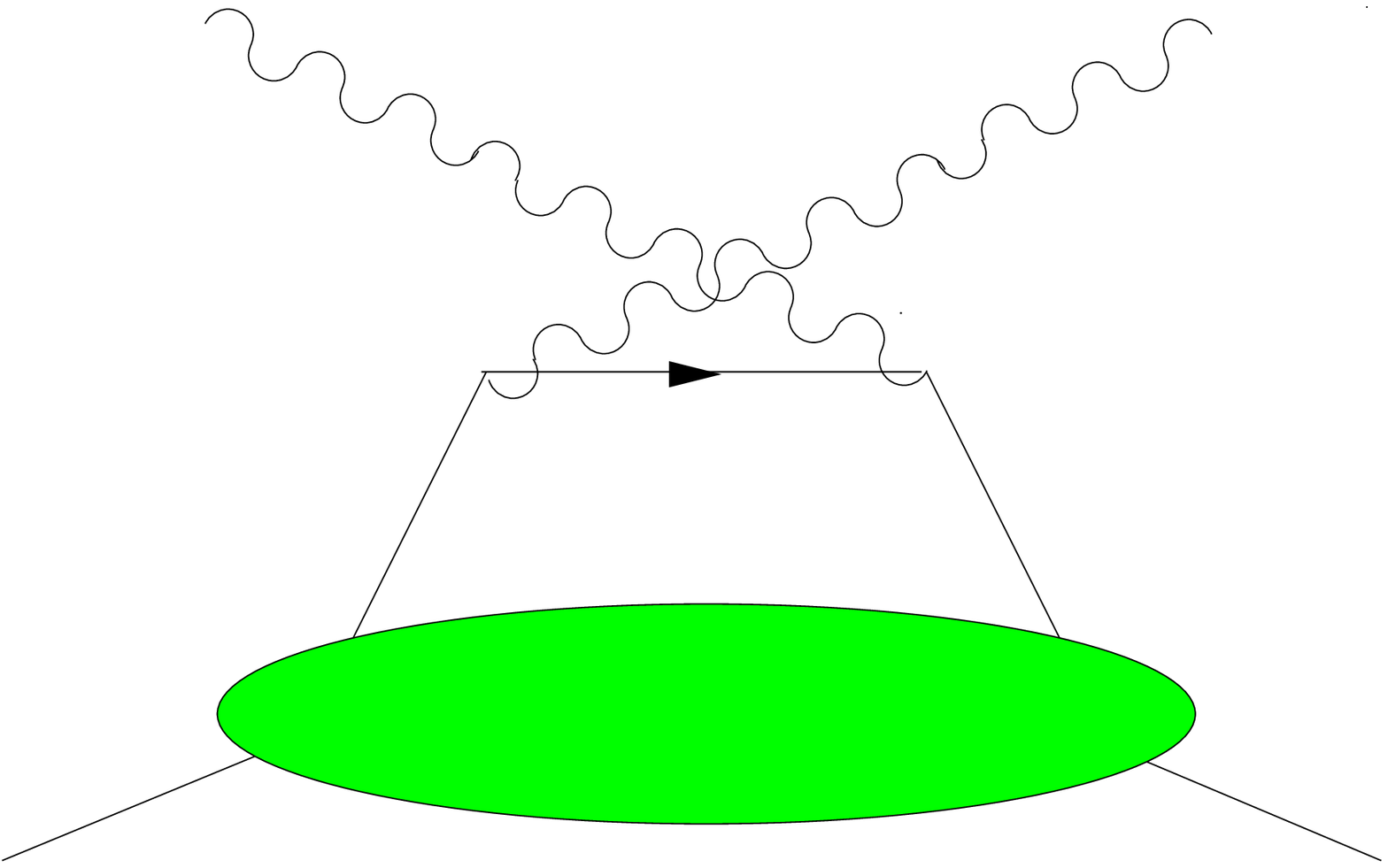}
\caption{Handbag diagrams for the Compton process in the scaling limit. The
plus-momentum fractions $x$, $\xi$, $\eta$ refer to the average proton
momentum $\frac{1}{2}(p+p')$.}
\label{haba}
\end{center}
\end{figure}

 In the region where the final photon virtuality is large, the amplitude is given by the
convolution of hard scattering coefficients, calculable in
perturbation theory, and generalized parton distributions, which
describe the nonperturbative physics of the process. To leading order
in $\alpha_s$ one then has the dominance of  the quark handbag diagrams of
Fig.~\ref{haba}.  
The analysis of these handbag diagrams  show the simple relations
\begin{eqnarray}
&& M^{\lambda'+,\lambda+} \Big|_{TCS}
  = \Big[ M^{\lambda'-,\lambda-} \Big]_{DVCS}^* , 
\nonumber \\
&&
M^{\lambda'-,\lambda-} \Big|_{TCS}
  = \Big[ M^{\lambda'+,\lambda+} \Big]_{DVCS}^*
\label{TCS-DVCS-amp}
\end{eqnarray}
between the helicity amplitudes for TCS and DVCS at equal values of
$\eta$ and $t$. For instance, 
\begin{eqnarray}
M^{+-,+-} \Big|_{TCS}
= \sqrt{1-\eta^2} ({\cal H}_1(-\eta,\eta,t) +\tilde{\cal
H}_1(-\eta,\eta,t) -\frac{\eta^2}{1-\eta^2}({\cal
E}_1(-\eta,\eta,t)+\tilde{\cal E}_1(-\eta,\eta,t) ))\;,
\label{TCS-amp}
\end{eqnarray}
where the  Compton form factors ${\cal H}_1,{\cal E}_1 ,\tilde{\cal H}_1,
 \tilde{\cal E}_1$
are defined as
\begin{eqnarray}
{\cal H}_1(\xi,\eta,t) &=& \sum_q e_q^2 \int_{-1}^{1} d x
\Big( \frac{1}{\xi-x-i\epsilon} - \frac{1}{\xi+x-i\epsilon} \Big)
H^q(x,\eta,t),
\nonumber \\
{\cal E}_1(\xi,\eta,t) &=& \sum_q e_q^2\int_{-1}^{1}d x
\Big( \frac{1}{\xi-x-i\epsilon} - \frac{1}{\xi+x-i\epsilon} \Big)
E^q(x,\eta,t),
\nonumber \\
\tilde{\cal H}_1(\xi,\eta,t) &=& \sum_q e_q^2 \int_{-1}^{1}d x
\Big( \frac{1}{\xi-x-i\epsilon} + \frac{1}{\xi+x-i\epsilon} \Big)
\tilde H^q(x,\eta,t),
\nonumber \\
\tilde{\cal E}_1(\xi,\eta,t) &=& \sum_q e_q^2 \int_{-1}^{1}d x
\Big( \frac{1}{\xi-x-i\epsilon} + \frac{1}{\xi+x-i\epsilon} \Big)\,.
\tilde E^q(x,\eta,t),
\label{htilde}
\end{eqnarray}
and $H^q(x,\eta,t),E^q(x,\eta,t), \tilde H^q(x,\eta,t), \tilde E^q(x,\eta,t)  $ are usual GPDs for a quark of flavour $q$ and electric charge $e\,e_q$.

 At this order the two processes carry the same information on the generalized 
quark distributions. This will not be true when higher order contributions are included but we will not 
consider these corrections here.

%

 The crucial ingredient to estimate the TCS amplitude at large energies  is a realistic model 
of GPDs at small skewedness.

\subsection{Modelizing GPDs}

We anticipate that singlet quark  GPDs give the dominant contributions to the TCS amplitude in that 
domain. Since gluon GPDs only enter the TCS amplitude at the $O(\alpha_{S})$ level, we feel justified 
in a first step to neglect their contributions and leave its study for a future work.
The  choice of factorization scale is a major issue since GPD evolution is particularly active in the small
$x$ domain. In this first study of the feasibility of the extraction of the TCS signal, we feel justified to
simplify our calculations by using  a factorization ansatz for the $t$ dependence of GPD's:

\begin{figure}
\begin{center}
\epsfxsize=0.39\textwidth
\epsffile{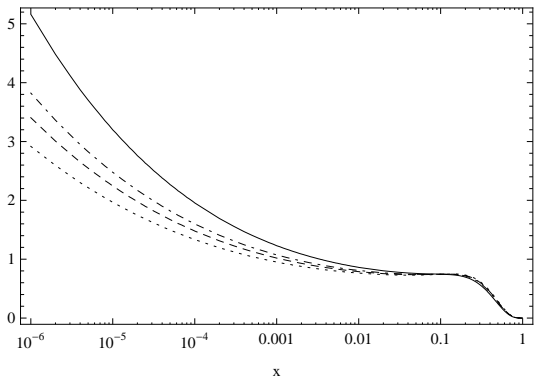}
\caption{
The NLO($\overline{MS}$) GRVGJR 2008 parametrization 
of $u(x) + \bar{u}(x)$ for different factorization scales 
$\mu_F^2 = 4$ 
(dotted)
, $5$ 
 (dashed)
, $6$ (dash-dotted)
, $10$ (solid) 
$\gev^2$.}
\label{xupl}
\end{center}
\end{figure}
\begin{figure}
\begin{center}
\epsfxsize=0.39\textwidth
     \epsffile{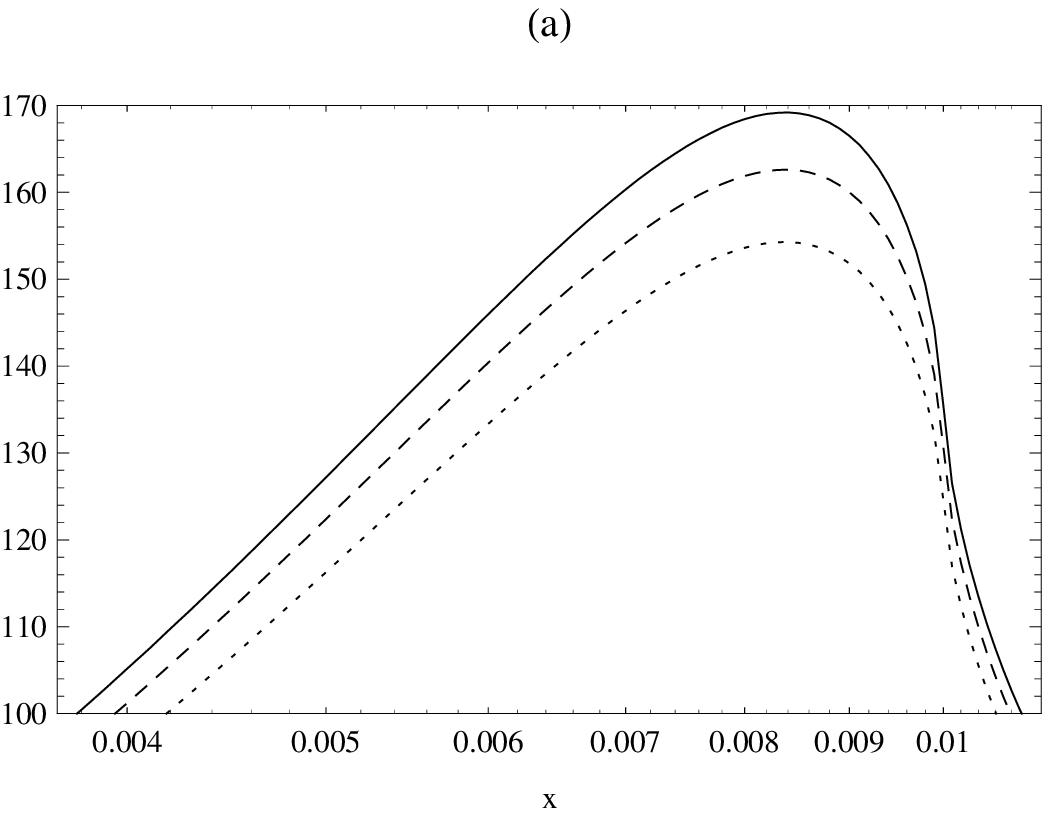}
\hspace{0.05\textwidth}
\epsfxsize=0.39\textwidth
\epsffile{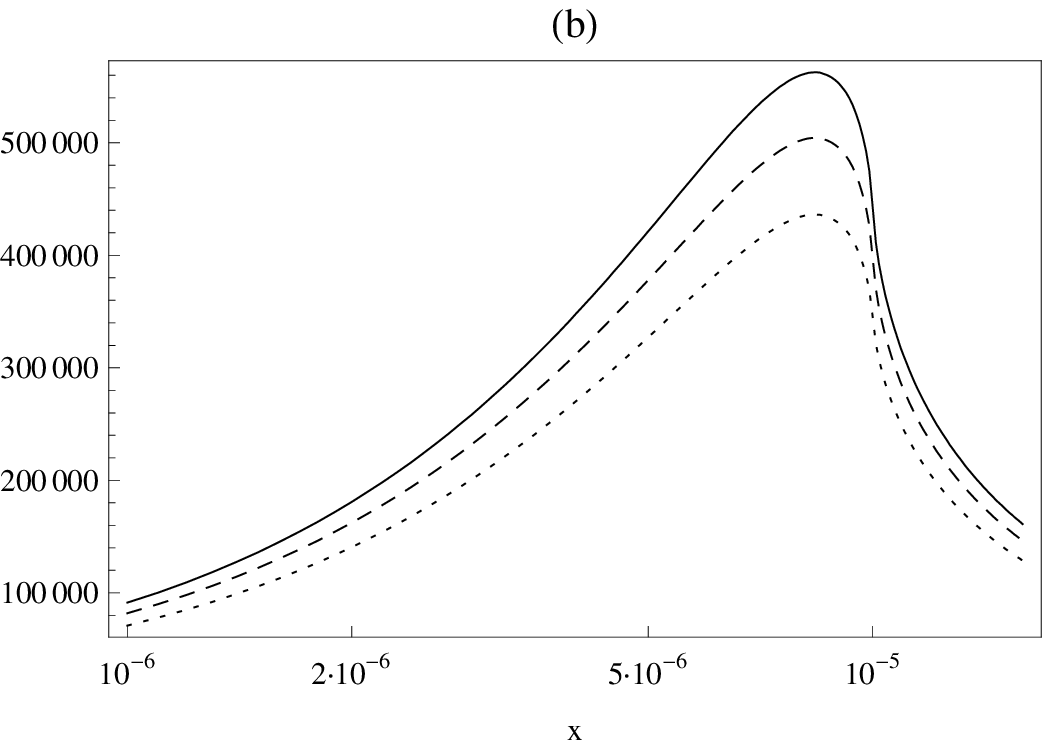}
\caption{$h_+^u(x,\eta)= h^u(x,\eta) - h^u(-x,\eta)$ for $\eta = 10^{-2}$ (a) 
and for $\eta =10^{-5}$ (b) for different factorization scales $\mu_F^2 = 4$ 
(dotted)
, $5$ 
 (dashed)
, $6$ (solid) 
$\gev^2$.} 
\label{hupl}
\end{center}
\end{figure}
\begin{eqnarray}
H^u(x,\eta,t) &=& h^u(x,\eta)\frac{1}{2}F_1^u(t) \nonumber \\
H^d(x,\eta,t) &=& h^d(x,\eta)F_1^d(t) \nonumber \\
H^s(x,\eta,t)&=& h^s(x, \eta)F_D(t) \nonumber
\end{eqnarray} 
and a double distribution ansatz for $h^q$ without any D-term:
\begin{eqnarray}
h^q(x,\eta) &=& \int_0^1 dx' \int _{-1+x'}^{1-x'} dy' 
\bigg[\delta(x-x'-\eta y') q(x') - \delta (x+x' -\eta y') \bar{q}(x')\bigg] \pi(x',y') 
\nonumber \\
\pi(x',y') &=& \frac{3}{4} \frac{(1-x')^2-y^{'2}}{(1-x')^3} \nonumber 
\end{eqnarray}
For the unpolarized distributions $q(x)$ and $\bar{q}(x)$ we 
take NLO($\overline{\rm MS}$) GRVGJR 2008 parametrization \cite{GRVGJR}. 
Their strong dependence of the factorization scale choice 
for small $x$ is shown on Fig.\ref{xupl}. This results in the strong dependence 
of $h^q$ for small values of $\eta$ as shown on Fig.\ref{hupl}.

\section{Cross section estimates}
Let us now estimate the 
different contributions to the lepton pair cross section for 
ultraperipheral collisions at the LHC.  Since the
cross sections decrease rapidly with $Q'^2$, we are interested in the kinematics of moderate $Q'^2$, 
say a few GeV$^2$, and large energy, thus very small values of $\eta$. 
Note however that for a given proton energy the photon flux is higher at 
smaller photon energy.

\subsection{The Bethe Heitler cross section}
\begin{figure}
\begin{center}
\epsfxsize=0.33\textwidth
\epsffile{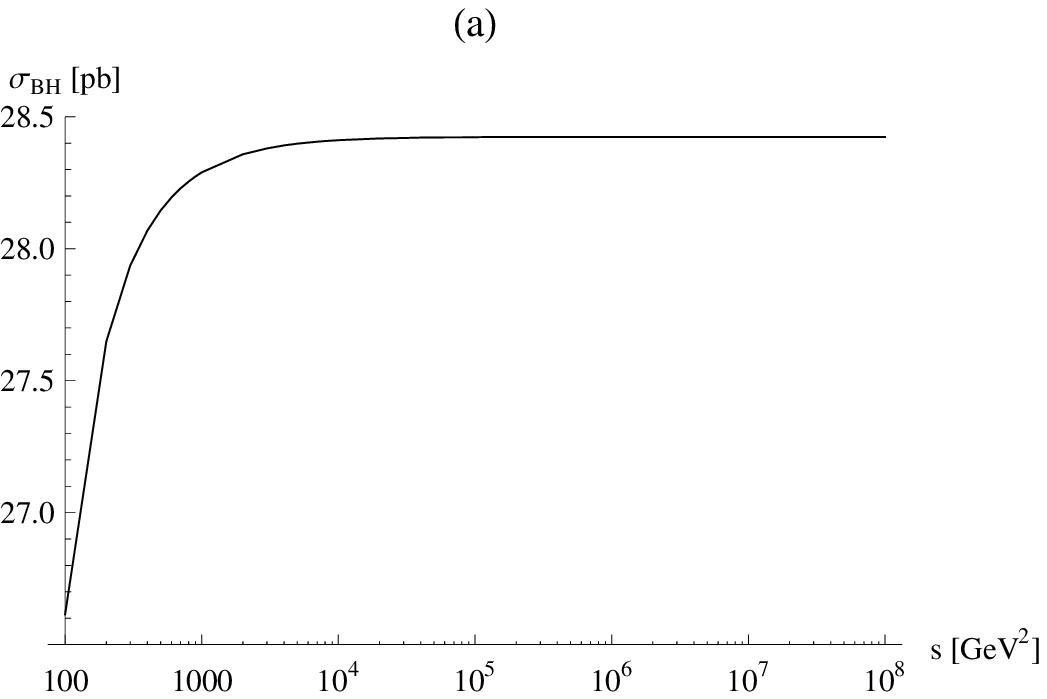}
\epsfxsize=0.33\textwidth
\epsffile{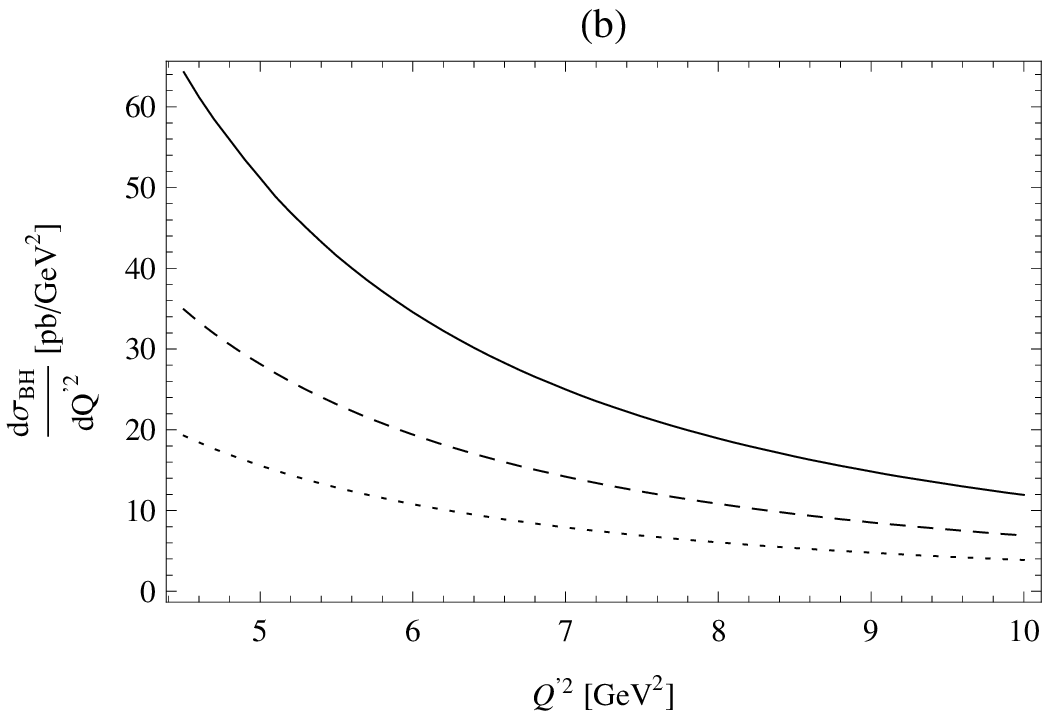}
\caption{(a) The BH cross section integrated over $\theta \in [\pi/4,3
\pi/4]$, $\varphi \in [0, 2\pi]$ , $Q'^2 \in [4.5,5.5]\gev^2$, $|t| \in [0.05,0.25] \gev^2$, as a function of $\gamma p$ c.m. energy squared $s$.
(b) The BH cross section integrated over $\varphi \in [0, 2\pi]$ , $|t| \in [0.05,0.25] \gev^2$, and various ranges of 
$\theta$ : $[\pi/3,2 \pi/3]$ (dotted), $[\pi/4,3 \pi/4]$ (dashed) 
and $[\pi/6,5 \pi/6]$ (solid), as a function of ${Q'}^2$ for $s=10^5 \gev^2$
} 
\label{BHs}
\end{center}
\end{figure}
The full Bethe Heitler cross section integrated over $\theta \in [\pi/4,3
\pi/4]$, $\varphi \in [0, 2\pi]$ , $Q'^2 \in [4.5,5.5]\gev^2$, $|t| \in [0.05,0.25] \gev^2$, as a function of $\gamma p$  energy squared $s$ is shown on Fig. \ref{BHs}a. We see that in the limit of large
 $s$ it is constant and equals $28.4 \pb$. 
On Fig. \ref{BHs}b, the Bethe Heitler contribution is shown as a function of $Q'^2$ when it is integrated over  $\varphi$ in the range $[0, 2 \pi]$, $-t $ in the range $[0.05, 0.25]$GeV$^2$ and for  $\theta$ integrated  in various ranges $[\pi/3,2 \pi/3]$, $[\pi/4,3 \pi/4]$ and $[\pi/6,5 \pi/6]$. 
As anticipated, the cross section grows much when small $\theta$ angles are allowed.
In the following we will use the limits $[\pi/4,3 \pi/4]$ where the cross section is sufficiently big and does not dominate too much over the Compton process.

\subsection{The TCS cross section}
\begin{figure}
\begin{center}
\epsfxsize=0.39\textwidth
\epsffile{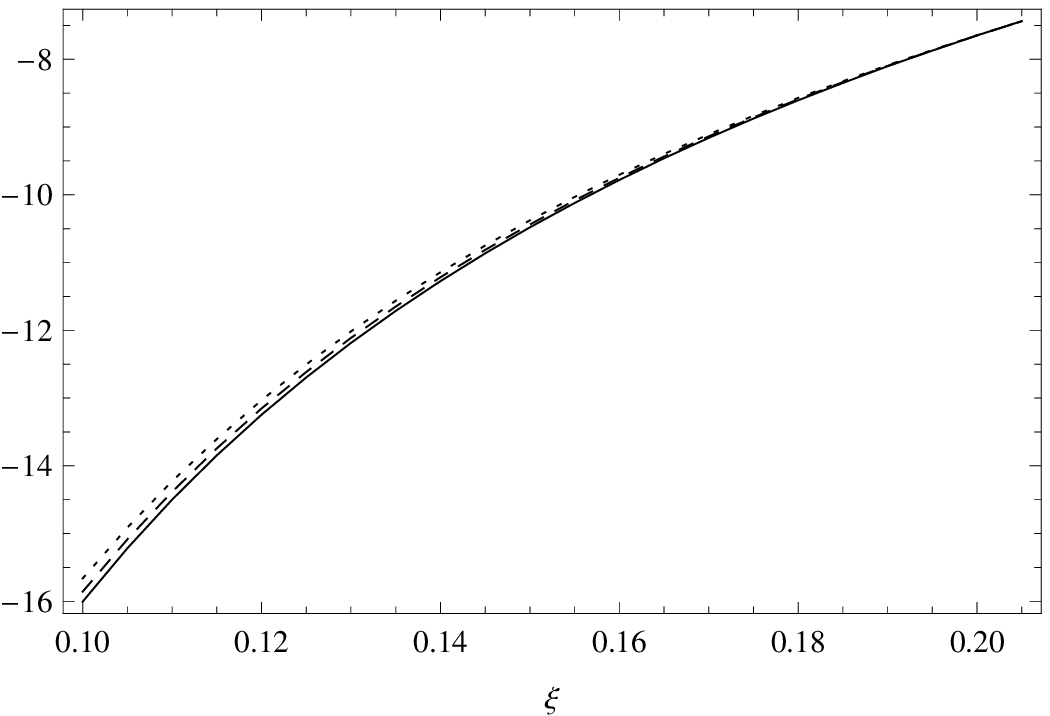}
\hspace{0.05\textwidth}
\epsfxsize=0.39\textwidth
\epsffile{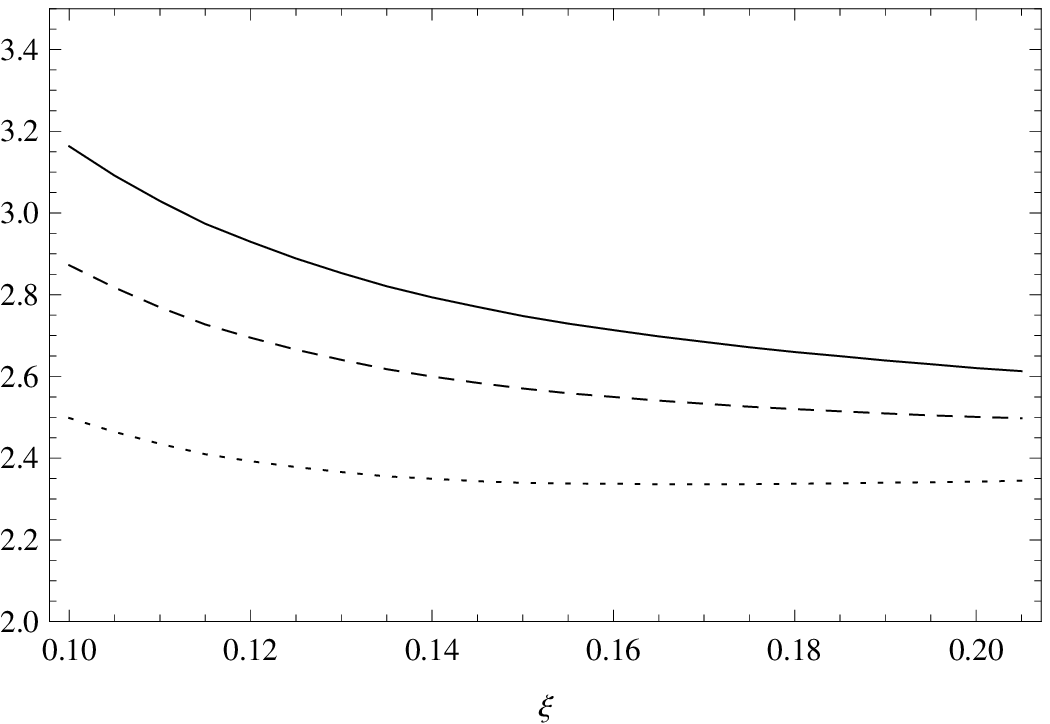}
\vspace{0.05\textwidth}
\epsfxsize=0.39\textwidth
\epsffile{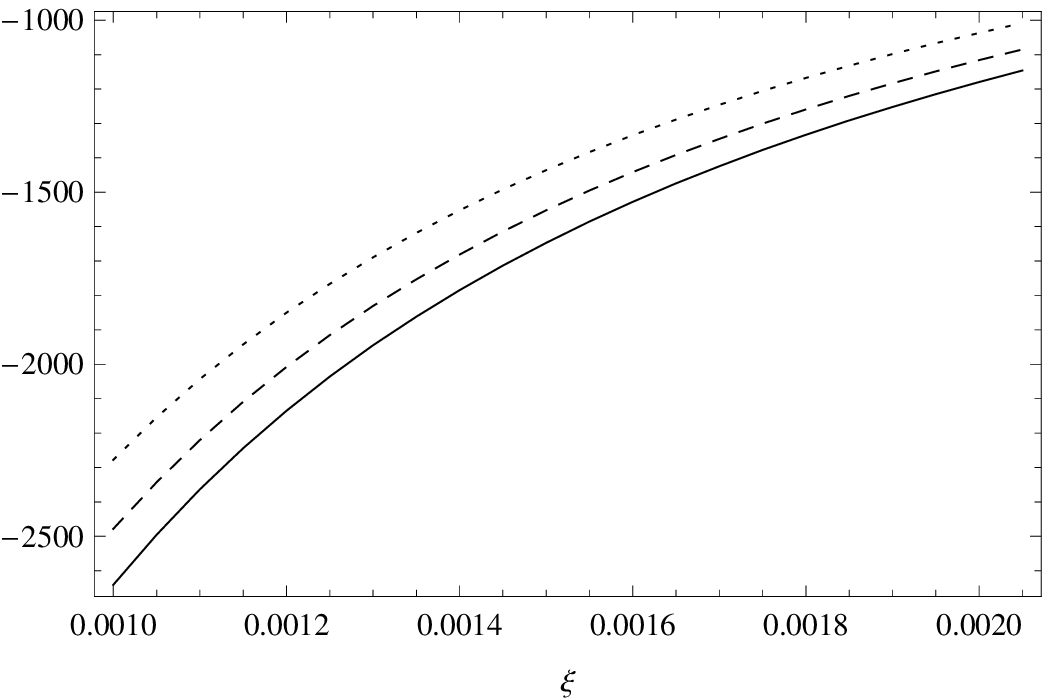}
\hspace{0.05\textwidth}
\epsfxsize=0.39\textwidth
\epsffile{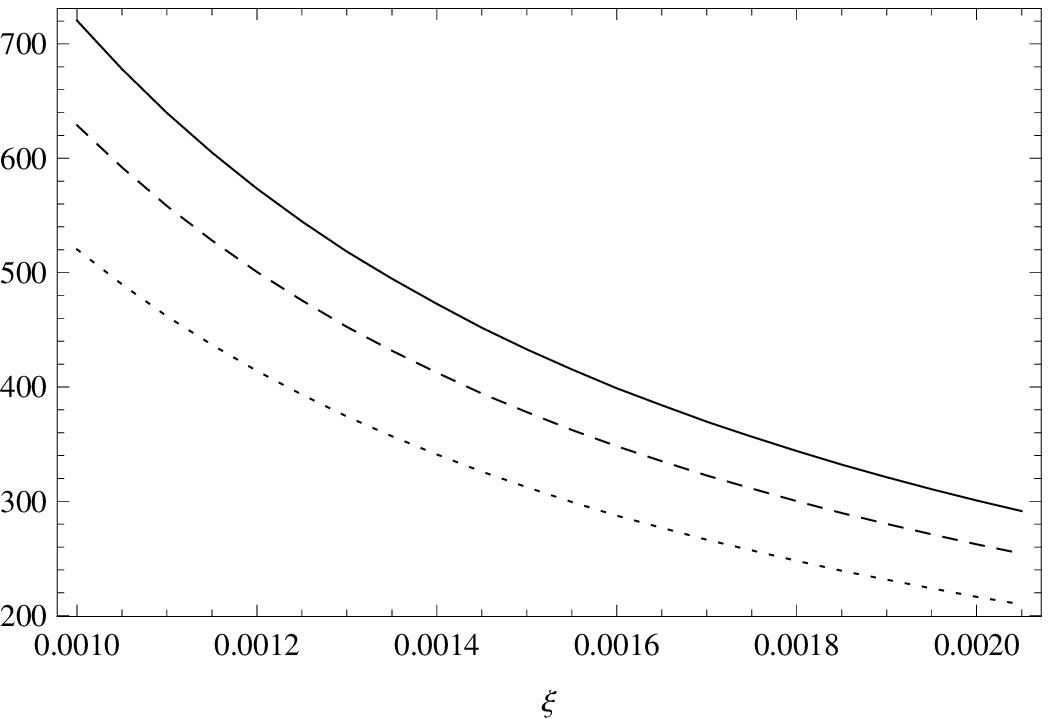}
\vspace{0.05\textwidth}
\epsfxsize=0.39\textwidth
\epsffile{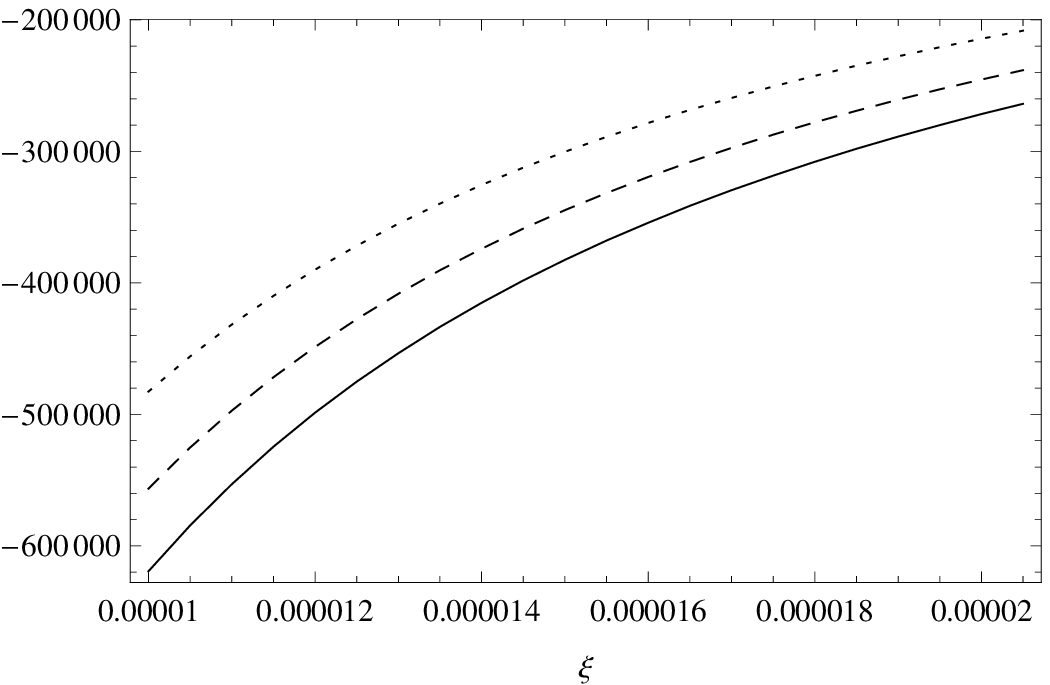}
\hspace{0.05\textwidth}
\epsfxsize=0.39\textwidth
\epsffile{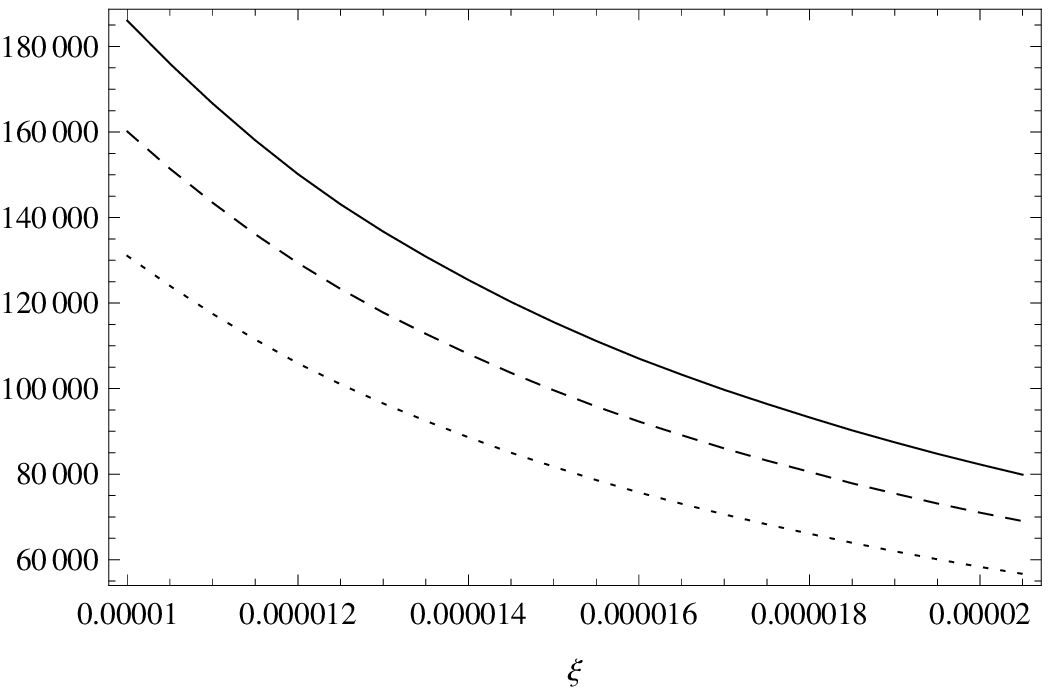}

\caption{$\im \mathcal{H}^u$ (left) and $\re \mathcal{H}^u$ (right) divided by $\frac{1}{2}F^u$ 
for various factorization scales $\mu_F^2 = 4$ (dotted), $5$ (dashed), $6$ (solid) $\gev^2$ 
and various ranges of $\xi$ : $[1\cdot 10^{-1},2\cdot 10^{-1}]$,$[1\cdot 10^{-3},2\cdot 10^{-3}]$,$[1\cdot 10^{-5},2\cdot 10^{-5}]$.}
\label{rehu_imhu}
\end{center}
\end{figure}

\begin{figure}
\begin{center}
\epsfxsize=0.39\textwidth
\epsffile{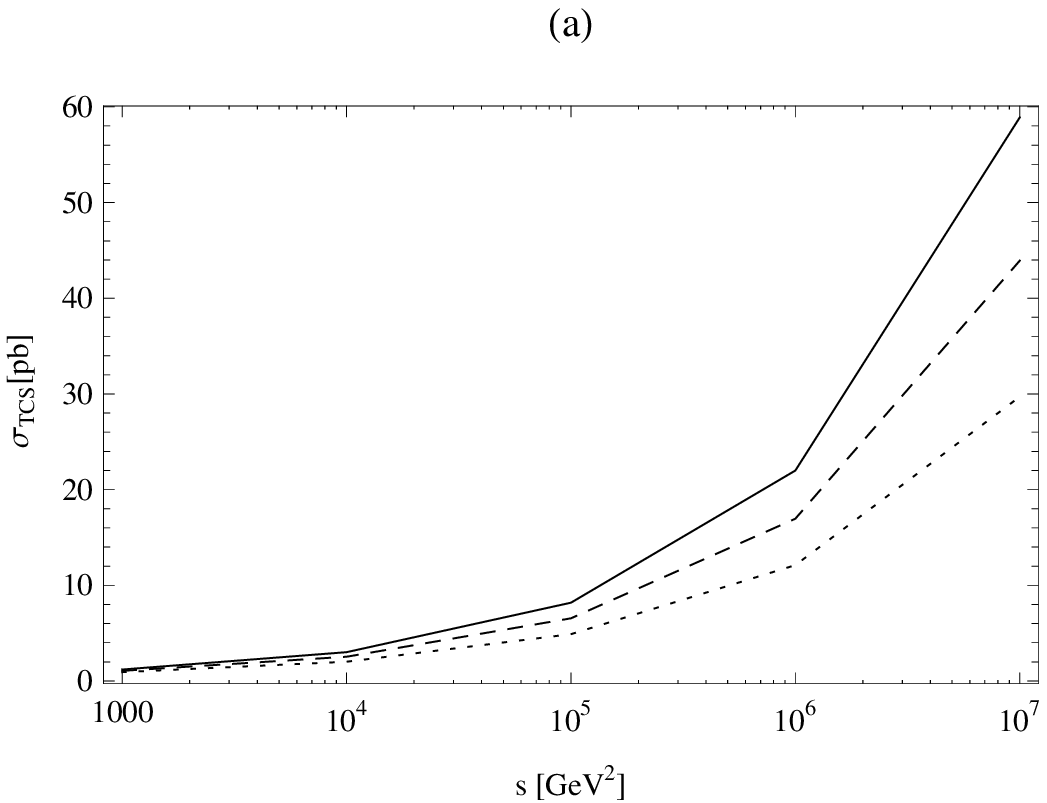}
\hspace{0.05\textwidth}
\epsfxsize=0.39\textwidth
\epsffile{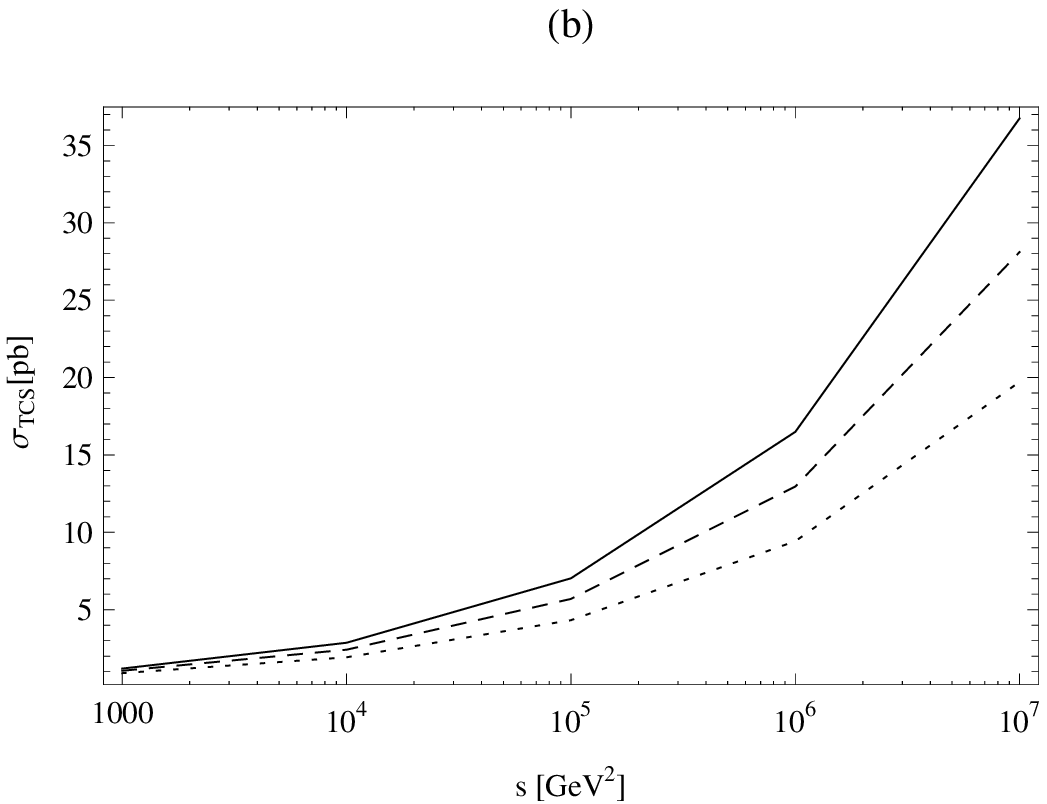}
\caption{$\sigma_{TCS}$ as a function of $\gamma p$ c.m. energy squared $s$, for 
GRVGJR2008 LO (a) and NLO (b) parametrizations, for different factorization scales 
$\mu_F^2 = 4$ (dotted), $5$ (dashed), $6$ (solid) $\gev^2$.}
\label{Sigma_TCS}
\end{center}
\end{figure}

Since the u-quark contribution of $H_1$ turns out to dominate
the TCS amplitude, we show in Fig. \ref{rehu_imhu} the real
 and imaginary part of $\mathcal{H}^u$ divided by $\frac{1}{2}F^u(t)$ for various factorization 
scales and various ranges of $\xi$. We observe that for small values of $\xi$, factorization scale 
dependence 
is quite strong. The same is seen on Fig. \ref{Sigma_TCS} where the full Compton cross section $\sigma_{TCS}$ is
 plotted as a function of the  photon-proton energy  squared $s$. For very high energies $\sigma_{TCS}$ calculated with $\mu_F^2 = 6 \gev^2$ is much bigger then with $\mu_F^2 = 4\gev^2$. 
Also predictions obtained using LO and NLO GRVGJR2008 PDFs differ significantly. 
\subsection{The interference cross section}

Since the amplitudes for the Compton and Bethe-Heitler
processes transform with opposite signs under reversal of the lepton
charge,  the interference term between TCS and BH is
odd under exchange of the $\ell^+$ and $\ell^-$ momenta.
It is thus possible to
 project out
the interference term through a clever use of
 the angular distribution of the lepton pair. 
The interference part of the cross-section for $\gamma p\to \ell^+\ell^-\, p$ with 
unpolarized protons and photons is given at leading order by
\begin{eqnarray}
   \label{intres}
\frac{d \sigma_{INT}}{d\qq^2\, dt\, d\cos\theta\, d\varphi}
= {}-
\frac{\alpha^3_{em}}{4\pi s^2}\, \frac{1}{-t}\, \frac{M}{Q'}\,
\frac{1}{\tau \sqrt{1-\tau}}\,
  \cos\varphi \frac{1+\cos^2\theta}{\sin\theta}
     \re\tilde{M}^{--} \; ,
\end{eqnarray}
with 
\begin{equation}
\label{mmimi}
\tilde{M}^{--} = \frac{2\sqrt{t_0-t}}{M}\, \frac{1-\eta}{1+\eta}\,
\left[ F_1 {\cal H}_1 - \eta (F_1+F_2)\, \tilde{\cal H}_1 -
\frac{t}{4M^2} \, F_2\, {\cal E}_1 \,\right],
\end{equation}
where $-t_0 = 4\eta^2 M^2 /(1-\eta^2)$ .
 With the integration limits symmetric about $\theta=\pi/2$ the interference
term is odd under $\varphi\to \pi+\varphi$ due to charge conjugation,
whereas the TCS and BH cross sections are even. One may thus extract the 
Compton amplitude through a study of
$\int\limits_0^{2\pi}d\phi\,\cos \phi \frac{d\sigma}{d\phi}$.

\begin{figure}
\begin{center}
\epsfxsize=0.39\textwidth
\epsffile{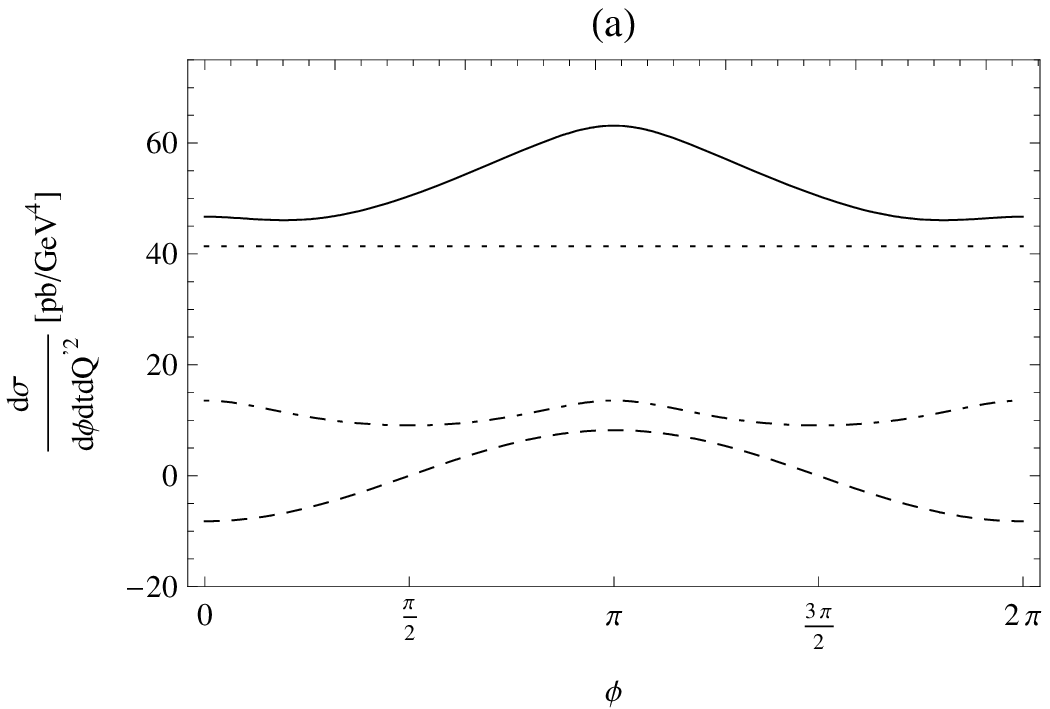}
\hspace{0.05\textwidth}
\epsfxsize=0.39\textwidth
\epsffile{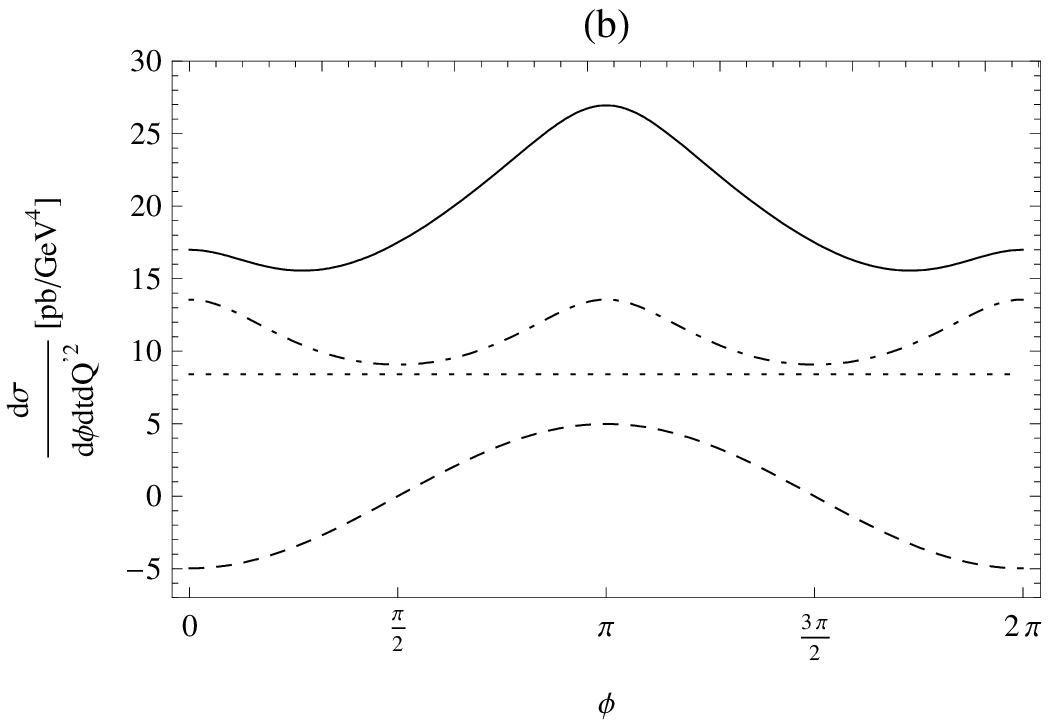}
\vspace{0.05\textwidth}
\epsfxsize=0.60\textwidth
\epsffile{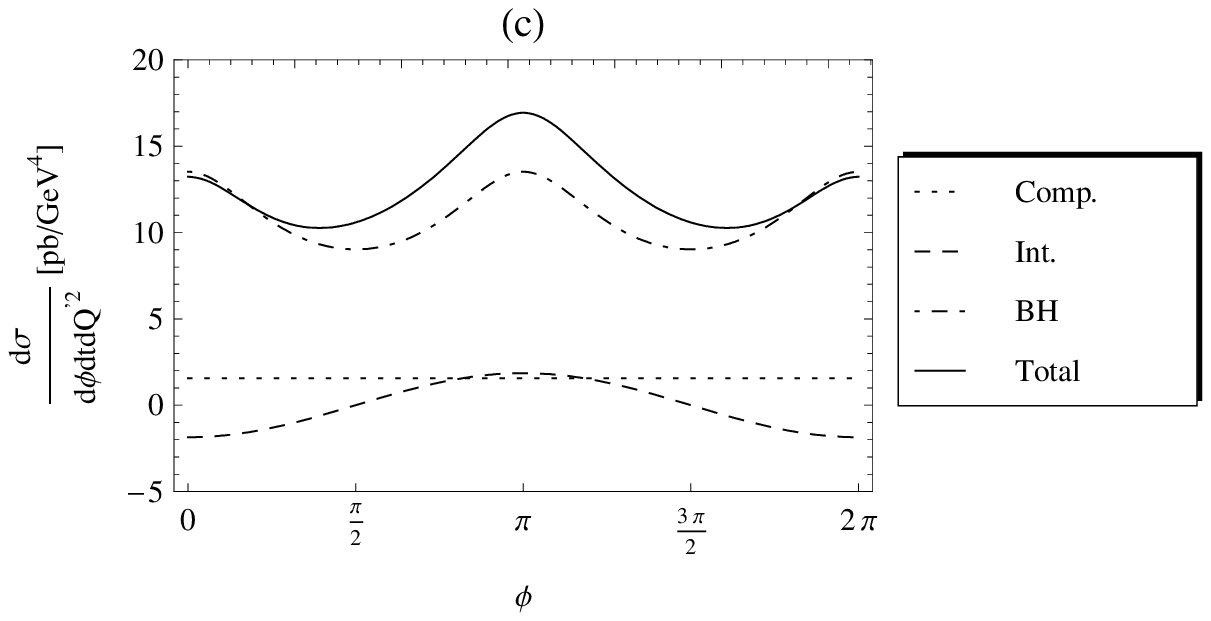}
\caption{
The differential cross sections (solid lines) for $t =-0.2 \gev^2$, ${Q'}^2 =5 \gev^2$ and integrated 
over $\theta = [\pi/4,3\pi/4]$, as a function of $\varphi$, for $s=10^7 \gev^2$ (a), 
$s=10^5 \gev^2$(b), $s=10^3 \gev^2$ (c) with $\mu_F^2 = 5 \gev^2$. We also display  the
Compton (dotted), Bethe-Heitler (dash-dotted) and Interference (dashed) contributions. 
}
\label{Interf}
\end{center}
\end{figure}

In Fig. \ref{Interf} we show the interference contribution to the cross section in comparison to the Bethe Heitler and Compton processes, for various values of photon proton  energy 
squared $s = 10^7 \gev^2,10^5 \gev^2,10^3 \gev^2$. We observe that for 
larger energies the  
Compton process dominates, whereas for $s=10^5 \gev^2$ all contributions are comparable. 
\subsection{Rate estimates}
As described in \cite{BeKlNy} the cross section for photoproduction in hadron collisions is given by:
\begin{equation}
\sigma_{pp}= 2 \int \frac{dn(k)}{dk} \sigma_{\gamma p}(k)dk
\end{equation}
where $\sigma_{\gamma p} (k)$ is the cross section for the 
$\gamma p \to pl^+l^-$ process and $k$ is the photon energy. 
$\frac{dn(k)}{dk}$ is an equivalent photon flux (the number of photons with energy $k$), and is given by \cite{DrZe}:
\begin{equation}
\frac{dn(k)}{dk} = \frac{\alpha}{2 \pi k} \left[1 + (1-\frac{2k}{\sqrt{s_{pp}}})^2\right]
\left(\ln A -\frac{11}{6} +\frac{3}{A} -\frac{3}{2A^2} +\frac{1}{3A^3} \right)
\end{equation}
where: $A =1 + \frac{0.71 \gev^2}{Q_{min}^2}$, $Q_{min}^2\approx \frac{4M_p^2k^2}{s_{pp}}$ is 
the minimal squared fourmomentum transfer for the reaction, and 
$s_{pp}$ is the proton-proton  energy squared ($\sqrt{s_{pp}} = 14 \tev$). 
The relationship between $\gamma p$  energy squared $s$ and k is given by:
\begin{equation}
s \approx 2\sqrt{s_{pp}}k \nonumber
\end{equation}
The pure Bethe - Heitler contribution to $\sigma_{p p}$, integrated over  $\theta = [\pi/4,3\pi/4]$, $\phi = [0,2\pi]$, $t =[-0.05 \gev^2,-0.25 \gev^2]$, ${Q'}^2 =[4.5 \gev^2,5.5 \gev^2]$, and photon energies $k =[20,900]\gev $  gives:
\begin{equation}
\sigma_{pp}^{BH} = 2.9 \pb \;.
\end{equation}  
The Compton contribution (calculated with NLO GRVGJR2008 PDFs, and $\mu_F^2 = 5 \gev^2$) gives:
\begin{equation}
\sigma_{pp}^{TCS} = 1.9 \pb\;. 
\end{equation}
We have choosen the range of photon energies in accordance with expected capabilities to tag photonenergies
at the LHC. This amounts to a large rate of order of $10^5$ events/year at the LHC with its nominal 
luminosity ($10^{34}\,$cm$^{-2}$s$^{-1}$). The rate remains sizeable for the lower luminosity which can be 
achieved in the first months of run. 
\section{Conclusion}

The operation of LHC as a heavy ion collider will enable us to study TCS on nuclei. Such scattering may 
occur in a
coherent way and its amplitude involves  nuclear GPDs \cite{NucGPD}. This  very interesting subject 
 definitely needs more work. Incoherent TCS which  occurs on quasi free neutrons and 
protons allows to study the GPDs of bound nucleons.
Cross sections are then roughly multiplied by $A^2$, where $A$ is the atomic number. One may thus expect 
sizeable rates in ion-ion collisions at the LHC.

In conclusion, let us stress that timelike Compton scattering in ultraperipheral collisions at hadron 
colliders opens a new way to measure generalized parton distributions. Our work has to be supplemented 
by studies of higher order contributions which  will involve the gluon GPDs; they will hopefully lead to
a weaker factorization scale dependence of the amplitudes.  

\section*{Acknowledgements} 
We acknowledge useful discussions and correspondence with 
I.V. Anikin, M. Diehl, D. M\"uller, M. Strikman,
O. Teryaev, F. Schwennsen and S. Wallon.
 This work is partly supported by the Polish Grant N202 249235,   
the French-Polish scientific agreement Polonium and by  
the ECO-NET program, contract 12584QK.



\begin{thebibliography}{99}


\bibitem{Muller:1994fv}
D.~M{\"u}ller {\em et al.},
Fortsch.\ Phys.\  {\bf 42}, 101 (1994);
X.~Ji,
Phys.\ Rev.\ Lett.\  {\bf 78}, 610 (1997);
A.~V.~Radyushkin,
Phys.\ Rev.\  {\bf D56}, 5524 (1997);
J.~C.~Collins and A.~Freund,
Phys.\ Rev.\  {\bf D59}, 074009 (1999).

\bibitem{gpdrev}
M.~Diehl,
  Phys.\ Rept.\  {\bf 388} (2003) 41;
  A.~V.~Belitsky and A.~V.~Radyushkin,
  Phys.\ Rept.\  {\bf 418}, 1 (2005);
S.~Boffi and B.~Pasquini,
Riv.\ Nuovo Cim. {\bf 30}, 387 (2007).



\bibitem{Burk}
M.~Burkardt,
  Phys.\ Rev.\  D {\bf 62}, 071503 (2000)
  and
  Int.\ J.\ Mod.\ Phys.\  A {\bf 18}, 173 (2003);
J.~P.~Ralston and B.~Pire,
  Phys.\ Rev.\  D {\bf 66}, 111501 (2002);
  M.~Diehl,
  Eur.\ Phys.\ J.\  C {\bf 25}, 223 (2002).

\bibitem{TCS}
E.~R.~Berger, M.~Diehl and B.~Pire,
  Eur.\ Phys.\ J.\  C {\bf 23}, 675 (2002).
  
  \bibitem{UPC}
K.~Hencken {\it et al.},
  Phys.\ Rept.\  {\bf 458}, 1 (2008).
G.~Baur, K.~Hencken, D.~Trautmann, S.~Sadovsky and Y.~Kharlov,
Phys.\ Rept.\ {\bf 364}, 359 (2002).



\bibitem{UPCLHC}
D.~d'Enterria, M.~Klasen and K.~Piotrzkowski,
Nucl.\ Phys.\ Proc.\ Suppl. B {\bf 179}, 1 (2008);
B.~Pire, L.~Szymanowski, F.~Schwennsen and S.~Wallon,
arXiv:0810.3817 [hep-ph].




  
  \bibitem{APT}
 I.~V.~Anikin, B.~Pire and O.~V.~Teryaev,
  Phys.\ Rev.\  D {\bf 62}, 071501 (2000);
    A.~V.~Belitsky and D.~M\"uller,
  Nucl.\ Phys.\  B {\bf 589} (2000) 611; 
  N.~Kivel, M.~V.~Polyakov, A.~Sch\"afer and O.~V.~Teryaev,
  Phys.\ Lett.\  B {\bf 497} (2001) 73; 
  A.~V.~Radyushkin and C.~Weiss,
  Phys.\ Rev.\  D {\bf 63} (2001) 114012;
  N.~Kivel, M.~V.~Polyakov and M.~Vanderhaeghen,
  Phys.\ Rev.\  D {\bf 63}, 114014 (2001).

  \bibitem{PSW}
  B.~Pire, L.~Szymanowski and J.~Wagner,
 Nucl.\ Phys.\ Proc.\ Suppl.\  {\bf 179-180}, 232 (2008).
  



\bibitem{GRVGJR}
  M.~Gluck, P.~Jimenez-Delgado and E.~Reya,
  Eur.\ Phys.\ J.\  C {\bf 53} (2008) 355.



 
\bibitem{BeKlNy} 
  C.~A.~Bertulani, S.~R.~Klein and J.~Nystrand,
  Ann.\ Rev.\ Nucl.\ Part.\ Sci.\  {\bf 55} (2005) 271

\bibitem{DrZe}
  M.~Drees and D.~Zeppenfeld,
  Phys.\ Rev.\  D {\bf 39} (1989) 2536.


\bibitem{NucGPD}
E. Berger  {\em et al.},  Phys.\ Rev.\ Lett.\  {\bf 87}, 142302 (2001);
 F.~Cano and B.~Pire,
  Eur.\ Phys.\ J.\  A {\bf 19}, 423 (2004);
  S.~Scopetta,
Phys.\ Rev. C {\bf 70}, 015205 (2004)
A.~Kirchner and D.~Mueller,
Eur.\ Phys.\ J.\ C {\bf 32}, 347 (2003);
V.~Guzey and M.~Strikman,
Phys.\ Rev.\ C {\bf 68}, 015204 (2003);
V.~Guzey,
arXiv:0801.3235 [nucl-th] and J.\ Phys.\ G {\bf 32}, 251 (2006);
V.~Guzey, A.~W.~Thomas and K.~Tsushima,
arXiv:0806.3288 [hep-ph];
A.~Freund and M.~Strikman,
Eur.\ Phys.\ J. C {\bf 33}, 53 (2004);
S.~Liuti and S.~K.~Taneja,
Phys.\ Rev. C {\bf 72}, 034902 (2005), {\it ibid}   
Phys.\ Rev. C {\bf 72}
032201 (2005).
  
  

\end{thebibliography}
\end{document}